\documentclass[aps,pra,reprint,10pt,superscriptaddress,floatfix]{revtex4-2}
\usepackage[utf8]{inputenc}
\usepackage[T1]{fontenc}
\usepackage{amsmath,amssymb,physics,bm}
\usepackage{microtype}
\usepackage{graphicx,booktabs}
\usepackage{orcidlink}
\usepackage{tikz}
\usepackage{pgfplots}
\usepackage{comment}
\pgfplotsset{compat=1.18}
\usepackage{hyperref}

\usepackage[english]{babel}

\graphicspath{{pdf/}}
\hypersetup{colorlinks=true,linkcolor=blue,citecolor=blue,filecolor=blue,urlcolor=blue,breaklinks=true}
\RequirePackage{color}

\begin{document}

\title{Source-driven torsional optical activity in geometrically chiral media}

\author{Edilberto O. Silva\,\orcidlink{0000-0002-0297-5747}}
\email{edilberto.silva@ufma.br}

\affiliation{Programa de P\'{o}s-Gradua\c{c}\~{a}o em F\'{i}sica, Universidade Federal do Maranh\~{a}o,\\ 65080-805, S\~{a}o Lu\'{i}s, Maranh\~{a}o, Brazil}

\affiliation{Coordena\c{c}\~ao do Curso de F\'{\i}sica -- Bacharelado, Universidade Federal do Maranh\~{a}o, 65085-580 S\~{a}o Lu\'{\i}s, Maranh\~{a}o, Brazil}

\begin{abstract}
We develop an effective field-theoretical model for source-driven electromagnetic waves in a geometrically chiral optical medium described by a uniform axial torsion. Starting from the gauge-invariant electromagnetic field strength, we introduce a Chern--Simons-type coupling between the vector
potential and the axial contortion of the medium, obtaining an inhomogeneous Maxwell system with prescribed external sources. For transverse cylindrical source profiles, we derive the coupled radial equations and solve them analytically for a circulating cylindrical-shell current. In the circular basis the homogeneous problem diagonalizes, producing torsion-split radial wavenumbers that become a longitudinal circular birefringence under guided-mode
boundary conditions. In the weak-torsion regime this birefringence produces a polarization rotation proportional to the propagation length and an oscillatory conversion between orthogonal linear polarizations. The source fixes the jump conditions and the relative excitation of the circular eigenmodes, leading to measurable modal-power and Poynting-flow signatures. A two-port scattering analysis shows that the apparent direction dependence remains Lorentz reciprocal, placing the effect in the class of reciprocal geometric optical activity rather than Faraday isolation.
\end{abstract}

\maketitle

\section{Introduction}
\label{sec:intro}

The interaction of light with matter is the cornerstone of modern optics.
Macroscopic phenomena such as refraction, absorption and optical activity are
dictated by the symmetry properties of the medium through which light
propagates.  Optical activity, rotation of the plane of linear
polarization, is traditionally traced to molecular or
crystallographic chirality, i.e. the absence of mirror symmetry in the
material's microstructure \cite{Barron2004,Tang2010}.  In recent years it has
become clear that chirality can also emerge from a deeper, geometric level:
non-Riemannian features of the effective spacetime that models the
medium \cite{Plebanski1960}.

Within the geometric (or gauge) theory of defects
\cite{Bilby1955,Kroener1980,Katanaev1992},
a continuous distribution of disclinations and dislocations is encoded in
spacetime curvature and torsion, respectively.
A uniform density of screw dislocations is therefore
geometrically equivalent to a spacetime endowed with constant axial torsion.
This "spiral-staircase'' geometry was first analyzed by Puntigam and
Hehl \cite{Puntigam1997} and later applied to condensed-matter contexts,
where it mimics a background effective magnetic field for
scalar or spinorial quasiparticles
\cite{Furtado2001,silvanetto2008,Bakke2012}.
When electromagnetic waves traverse such a torsional background,
Maxwell's equations admit two circularly polarized eigenmodes with different phase velocities, giving rise to geometric circular birefringence
\cite{Obukhov2003,Itin2010}.  To date, however, these analyses have
been restricted to the source-free case ($J^\mu=0$); they describe how
light propagates in a torsional medium but not how it is generated, driven or manipulated by currents inside such a chiral background.

In this paper we remove that limitation.
Starting from an effective action for electrodynamics in a fixed torsional background, we incorporate a prescribed external current $J^\mu$ and then
specialize to transverse cylindrical source profiles.  We then solve them analytically for a canonical azimuthal shell current, an idealized yet experimentally relevant configuration that emulates a thin, current-carrying coil.  The interplay between the external source and the uniform torsion produces three central results.  First, the torsion term mixes the radial and azimuthal field amplitudes, and the circular combinations $a_\pm=a_\rho\pm i a_\phi$ become the natural eigenmodes.  Second, the two circular components acquire different effective transverse wavenumbers, which translate into distinct longitudinal propagation constants once a guided-mode boundary condition is imposed; in the weak-torsion limit this gives a polarization rotation $\Delta\theta\simeq \Omega L$ and a conversion efficiency $\eta_{x\to y}(L)=\sin^2(\Omega L)$.  Third, reversing the propagation direction changes the sign of the $k_z\Omega$ coupling.  This leads to an apparent direction-dependent Jones response, but the two-port scattering analysis shows that torsion alone remains Lorentz reciprocal.  We therefore distinguish the resulting geometric optical activity from genuine isolator-grade non-reciprocity, which requires additional time-reversal-symmetry breaking or device-level ingredients \cite{Yu2009,Raghu2008}.

The advance over previous source-free analyses is therefore not simply the inclusion of a nonzero right-hand side.  In the present formulation the source determines how the torsional eigenmodes are excited, fixes the matching conditions at the current shell, and allows one to compute observable Jones, Mueller, and Stokes responses for a driven optical field.  Thus the formalism connects the geometric birefringence of a torsional background with experimentally accessible polarization-generation and conversion protocols.

\begin{table*}[t]
\caption{Conceptual distinction between source-free torsional optics and the source-driven problem addressed here.}
\label{tab:comparison}
\begin{ruledtabular}
\begin{tabular}{lll}
 & Source-free models & Present model \\
\hline
Current & $J^\mu=0$ & prescribed $J^\mu\neq0$ \\
Main object & propagation eigenmodes & driven excitation and matching \\
Observable & circular birefringence & Jones/Mueller conversion response \\
Boundary data & homogeneous fields & source-induced jump conditions \\
\end{tabular}
\end{ruledtabular}
\end{table*}

Because the torsion parameter $\Omega$ is set by the density of screw dislocations, these results suggest a geometric avenue for designing metamaterials whose optical response can be tuned by controlled defect engineering \cite{Pendry2006,Kildishev2013}.  Beyond photonics, our formalism offers a blueprint for probing torsion in topological semimetals \cite{Cortijo2015,Pikulin2016}, where lattice dislocations act as sources of emergent geometry.  Taken together, the present study establishes uniform torsion as a versatile, purely geometric resource for controlling the polarization of light and for engineering direction-dependent optical responses.

\section{Background geometry}

We model the medium as a Riemann--Cartan defect geometry with vanishing curvature and non-vanishing torsion produced by a continuous density of parallel screw dislocations.  Locally, this geometry is represented by a spiral-staircase tetrad and reduces to the Minkowski geometry when the defect density is removed.  In cylindrical coordinates $x^{\mu}=(t,\rho,\phi,z)$ and metric signature $(-,+,+,+)$ the line element is
\begin{equation}
ds^{2}= -c^{2}\,dt^{2}
        + d\rho^{2}
        + \rho^{2}\,d\phi^{2}
        +\bigl(dz+\Omega\,\rho^{2}d\phi\bigr)^{2},
\label{eq:metric_spacetime}
\end{equation}
where the torsion parameter $\Omega \equiv b\sigma/2$
encodes a surface density $\sigma$ of screw dislocations with Burgers vector magnitude~$b$ \cite{Katanaev1992,silvanetto2008}.
The mixed term $dz+\Omega\rho^{2}d\phi$ is the hallmark of the
\emph{spiral solution}: it twists each transverse plane by an amount
proportional to $\rho^{2}$, and it disappears in the defect-free limit
$\Omega\!\to\!0$, yielding ordinary Minkowski space. We keep $c$ explicit throughout; in natural units, one sets $c=1$.

All torsional information is carried by a single non-vanishing component of the torsion tensor,
\begin{equation}
T^{z}{}_{\phi\rho}=-T^{z}{}_{\rho\phi}=2\,\Omega\rho,
\label{eq:torsion_component}
\end{equation}
which in turn produces the contortion element
$K_{\rho\phi z}=\Omega\rho$ (see Appendix~\ref{app:tetrad} for the
projection into an orthonormal frame).  In the effective axial Maxwell theory developed below, this contortion generates the chiral coupling
$K^{\nu}{}_{\mu\lambda}F^{\mu\lambda}$,
ultimately responsible for the circular birefringence and the geometric
polarization rotation analyzed in the following sections.

\section{Covariant Electrodynamics with Torsion and Sources}\label{sec:maxwell_sources}

To investigate the influence of torsion in the presence of sources, we begin with the general covariant formulation of Maxwell's equations. We consider a Riemann--Cartan spacetime, which is a manifold equipped with both a metric tensor, $g_{\mu\nu}$, and an independent affine connection, $\Gamma^\lambda_{\mu\nu}$. The antisymmetric part of the connection defines the torsion tensor, $T^\lambda{}_{\mu\nu} = \Gamma^\lambda_{[\mu\nu]}$. The connection can be decomposed as
\begin{equation}
    \Gamma^\lambda_{\mu\nu} = \{^{\,\lambda}_{\mu\nu}\} + K^\lambda{}_{\mu\nu},
    \label{eq:connection_decomposition}
\end{equation}
where $\{^{\,\lambda}_{\mu\nu}\}$ are the Christoffel symbols, determined by the metric $g_{\mu\nu}$, and $K^\lambda{}_{\mu\nu}$ is the contortion tensor, which is constructed from the torsion tensor. This decomposition separates the familiar Riemannian geometry from the effects of torsion.

The electromagnetic field strength tensor, $F_{\mu\nu}$, is defined via the exterior derivative of the four-potential $A_\mu$:
\begin{equation}
    F_{\mu\nu} = \partial_\mu A_\nu - \partial_\nu A_\mu.
    \label{eq:field_tensor_def}
\end{equation}
This choice is deliberate. We do not replace the exterior derivative by a torsionful covariant curl in the definition of $F_{\mu\nu}$; doing so would generally obscure the standard $U(1)$ gauge invariance of the electromagnetic field. The homogeneous Maxwell identity is therefore the exterior-form identity
\begin{equation}
    \mathrm{d}F=0,
    \qquad
    \partial_\lambda F_{\mu\nu}
    +\partial_\mu F_{\nu\lambda}
    +\partial_\nu F_{\lambda\mu}=0 .
    \label{eq:maxwell_hom}
\end{equation}
No torsionful covariant derivative is involved in Eq.~\eqref{eq:maxwell_hom}.
This point is important because the electromagnetic Bianchi identity follows
from $F=\mathrm{d}A$ and remains connection independent.

The standard minimally coupled Maxwell action in a Riemann--Cartan
background, with the same definition $F=\mathrm{d}A$, does not by itself
generate a direct torsion--field coupling. In the present work, torsion is
therefore introduced as an effective material response of the structured
chiral medium. The additional term can be motivated by the tensor structure
obtained from a trace-free axial contortion in the covariant divergence of
$F^{\mu\nu}$, but the field equation used below is defined most cleanly
through a gauge-compatible effective action.

\subsection{Effective axial action and field equation}
\label{subsec:action_principle}

For a uniform screw-dislocation background, the relevant part of the contortion is its axial dual. To avoid any ambiguity between Levi--Civita symbols and tensor densities, we define the contravariant Levi--Civita tensor by
\begin{equation}
    E^{\alpha\nu\mu\lambda}\equiv
    \frac{\epsilon^{\alpha\nu\mu\lambda}}{\sqrt{-g}},
    \qquad \epsilon^{t\rho\phi z}=+1,
    \label{eq:levicivita_tensor_def}
\end{equation}
and lower indices with the metric. The axial pseudovector $\Theta^\alpha$ is introduced through
\begin{equation}
    K^\nu{}_{\mu\lambda}
    =\frac{1}{2}E^\nu{}_{\mu\lambda\alpha}\Theta^\alpha .
    \label{eq:K_dual_theta}
\end{equation}
In the local orthonormal frame adapted to the spiral-staircase geometry, the axial vector is chosen to point along the screw-dislocation axis. With the conventions used in the transverse equations below, its only nonzero physical component is
\begin{equation}
    \Theta^{\hat z}=2\Omega,
    \label{eq:theta_omega_identification}
\end{equation}
so that the dual-contortion term reproduces the transverse couplings $\pm 2ik_z\Omega$ in Eq.~\eqref{eq:coupled_system}. This equation ensures the normalization of $\Theta^\alpha$ relative to the torsion parameter that appears in the defect metric. Both $\Theta^{\hat z}$ and $\Omega$ have dimensions of inverse length. Changing the orientation convention reverses the sign assigned to $\Omega$, but all observable rotation angles depend on the product $\Omega L$ together with the chosen handedness convention.

The lowest-order effective action that couples the electromagnetic field to this axial geometrical background is the Carroll--Field--Jackiw/Chern--Simons-type action \cite{Carroll1990,Wilczek1987,Qi2008,Essin2009}
\begin{align}
&S[A]=\int d^4x\, \sqrt{-g}\notag\\ &\times \left[
-\frac{1}{4}F_{\mu\nu}F^{\mu\nu}
+\frac{1}{4}\Theta_\alpha
 E^{\alpha\nu\mu\lambda}A_\nu F_{\mu\lambda}
-\mu_0 A_\nu J^\nu
\right].
\label{eq:effective_action}
\end{align}
Here $E^{\alpha\nu\mu\lambda}$ is a true tensor, not a density; therefore, the factor $\sqrt{-g}$ in the integration measure is the only metric density appearing in the action. The sign convention in Eq.~\eqref{eq:effective_action} is chosen so that the axial term contributes with the same sign as the contortion term in Eq.~\eqref{eq:maxwell_sources_torsion} below. For constant $\Theta^\alpha$, the Chern--Simons density changes under $A_\mu\to A_\mu+\partial_\mu\chi$ by a boundary term, provided the Bianchi identity~\eqref{eq:maxwell_hom} holds. Thus, the effective model preserves the usual gauge structure under the assumptions used here.

Varying Eq.~\eqref{eq:effective_action} with respect to $A_\nu$ and discarding boundary terms gives
\begin{align}
&\delta S=\int d^4x\, \sqrt{-g}\notag\\&\times \left[
\frac{1}{\sqrt{-g}}\partial_\mu\!\left(\sqrt{-g}F^{\mu\nu}\right)
+\frac{1}{2}\Theta_\alpha E^{\alpha\nu\mu\lambda}F_{\mu\lambda}
-\mu_0J^\nu
\right]\delta A_\nu .
\label{eq:variation_effective_action}
\end{align}
Since $\delta A_\nu$ is arbitrary, the Euler--Lagrange equation is
\begin{equation}
\frac{1}{\sqrt{-g}}\partial_\mu\!\left(\sqrt{-g}F^{\mu\nu}\right)
+\frac{1}{2}\Theta_\alpha E^{\alpha\nu\mu\lambda}F_{\mu\lambda}
=\mu_0J^\nu .
\label{eq:eom_theta_cov}
\end{equation}
Using the duality relation~\eqref{eq:K_dual_theta}, this can be written as
\begin{equation}
    \frac{1}{\sqrt{-g}}\partial_\mu\!\left(\sqrt{-g}F^{\mu\nu}\right)
    +K^\nu{}_{\mu\lambda}F^{\mu\lambda}
    =\mu_0J^\nu .
    \label{eq:maxwell_sources_torsion_cov}
\end{equation}
In the following sections, the equations are projected onto the local orthonormal cylindrical frame. In that frame, the Maxwell divergence produces the standard radial operator $\partial_\rho^2+(1/\rho)\partial_\rho$, while the axial term becomes the simple algebraic mixing displayed in Eq.~\eqref{eq:coupled_system}. For compactness, and only after projection onto the orthonormal cylindrical basis, we use the schematic notation
\begin{equation}
    \mathcal{D}_\mu F^{\mu\nu}
    +K^\nu{}_{\mu\lambda}F^{\mu\lambda}
    =\mu_0J^\nu,
    \label{eq:maxwell_sources_torsion}
\end{equation}
where $\mathcal{D}_\mu$ denotes the projected Maxwell divergence, including the cylindrical measure and tetrad factors. This schematic projected equation is the effective inhomogeneous Maxwell equation adopted in the rest of the paper. In this formulation, the torsional contribution is not an artifact of replacing partial derivatives by a torsionful covariant derivative. It is a non-minimal optical response induced by the axial geometrical chirality of the medium.

Taking the ordinary divergence of Eq.~\eqref{eq:eom_theta_cov} gives the compatibility condition for charge conservation. The Maxwell contribution has vanishing ordinary divergence identically when written in density form, because it is the divergence of an antisymmetric tensor density, while the Chern--Simons term vanishes for constant $\Theta^\alpha$ by the Bianchi identity. Therefore
\begin{equation}
    \partial_\nu J^\nu=0
    \label{eq:charge_conservation}
\end{equation}
follows consistently for the prescribed external current. Equivalently, the torsion term may be moved to the right-hand side and interpreted as a geometry-induced polarization current,
\begin{equation}
    \mu_0 J^\nu_{\rm eff}
    =\mu_0J^\nu-K^\nu{}_{\mu\lambda}F^{\mu\lambda}.
    \label{eq:effective_current}
\end{equation}
For the stationary shell source studied below, the external current is divergence-free in the distributional sense: it is purely azimuthal, has no endpoints, and carries no net charge accumulation. Equation~\eqref{eq:maxwell_sources_torsion} therefore provides a gauge-compatible effective description of how the screw-dislocation geometry modifies the electromagnetic response. The torsion term breaks the degeneracy between the two circular components and gives rise to the chiral phenomena explored in the following sections.

\section{General Analytical Formalism in a Torsional Background}
\label{sec:general_formalism}

To obtain concrete physical predictions, we now apply the general theory from Sec.~\ref{sec:maxwell_sources} to a specific, highly symmetric configuration. We adopt cylindrical coordinates $(t, \rho, \phi, z)$ and consider a spacetime background where the only non-Riemannian feature is a uniform, constant torsion directed along the $z$-axis. Such a background, often called the "spiral staircase" or "screw dislocation" spacetime, is the simplest model for a medium with uniform geometric chirality.

The non-zero components of the contortion tensor are most easily expressed in a local orthonormal frame (or tetrad), which is necessary to correctly project tensor components into physically measurable quantities. As detailed in Appendix \ref{app:tetrad}, the only non-vanishing physical components are:
\begin{equation}
    K^{\hat\rho}{}_{\hat z\hat\phi}
    =
    -K^{\hat\phi}{}_{\hat z\hat\rho}
    =
    \Omega ,
    \label{eq:contortion_components}
\end{equation}
where $\Omega$ is the torsion parameter, having dimensions of inverse length, which quantifies the density of screw dislocations. The hats ($\hat{\cdot}$) denote the orthonormal frame vectors.

We seek wavelike solutions propagating along the $z$-axis, so we employ an ansatz for the four-potential $A_\mu$ that separates the radial dependence from the other coordinates:
\begin{equation}
    A_\mu(x) = a_\mu(\rho)\, e^{i (k_z z - \omega t + m\phi)}.
    \label{eq:ansatz_with_m}
\end{equation}
Here, $a_\mu(\rho)$ are the radial amplitude functions, $\omega$ is the angular frequency, $k_z$ is the longitudinal wavenumber, and $m \in \mathbb{Z}$ is the azimuthal mode number, which quantifies the orbital angular momentum of the wave.

We restrict attention to the transverse, source-driven sector relevant for the circulating shell current considered below. In the temporal gauge, we set $A_t=0$ and impose the transverse modal constraint on the two physical components. For the neutral azimuthal source used here, $J^t=J^z=0$, and we consistently take $A_z=0$ within this idealized transverse block. Thus, Eqs.~\eqref{eq:coupled_system} should be read as the closed transverse sector of the effective theory, not as the most general vector solution. The transverse reduction is therefore a controlled modal truncation rather than a statement about the complete electromagnetic spectrum of a cylindrical waveguide. Longitudinal, charge-density, or hybrid waveguide sources would require solving the scalar and axial amplitudes together with the transverse sector.

A note on the azimuthal mode number is in order.
The source in Eq.~\eqref{eq:current_profile} has no $\phi$ dependence;
its Fourier decomposition in $m$ therefore contains only the $m=0$
component. Accordingly, we set $m=0$ in the numerical evaluation of
Sec.~\ref{sec:numerical_results}. This choice carries a further theoretical advantage: for $m=0$, the spin-connection (vector-Laplacian) cross-coupling terms of the full cylindrical vector wave equation, which are proportional to $\pm 2im/\rho^{2}$, vanish
identically. The scalar-envelope approximation underlying
Eq.~\eqref{eq:coupled_system} is therefore \emph{exact} for $m=0$
within the transverse truncation, without any additional smallness
assumption on $\Omega$ or $R$. The general-$m$ case introduces
geometric cross-coupling corrections of relative order
$\xi\equiv m/(k_z\Omega R^2)$, which are discussed in
Appendix~\ref{app:axial_projection}.

Our task is to substitute this transverse ansatz into the governing equation~\eqref{eq:maxwell_sources_torsion}. Expressing the fields and currents in cylindrical coordinates and projecting the axial term with the conventions summarized in Appendix~\ref{app:axial_projection}, we obtain a coupled system of ordinary differential equations for the transverse components of the potential, $a_\rho(\rho)$ and $a_\phi(\rho)$:
\begin{subequations}
\label{eq:coupled_system}
\begin{align}
    &\left[ \frac{\dd^2}{\dd\rho^2} + \frac{1}{\rho} \frac{\dd}{\dd\rho} - \left( \frac{m^2}{\rho^2} + \gamma^2 \right) \right] a_\rho(\rho) \nonumber\\
    &\hspace{4cm} + 2i k_z \Omega a_\phi(\rho) = \mu_0 J^\rho(\rho),
    \label{eq:wave_arho_final} \\
    &\left[ \frac{\dd^2}{\dd\rho^2} + \frac{1}{\rho} \frac{\dd}{\dd\rho} - \left( \frac{m^2}{\rho^2} + \gamma^2 \right) \right] a_\phi(\rho) \nonumber\\
    &\hspace{4cm} - 2i k_z \Omega a_\rho(\rho) = \mu_0 J^\phi(\rho),
    \label{eq:wave_aphi_final}
\end{align}
\end{subequations}
where we have defined the parameter $\gamma^2 \equiv k_z^2 - \omega^2/c^2$. The sign of $\gamma^2$ determines the radial character of the mode. For $\gamma^2>0$, the radial dependence is evanescent, and the solutions are most naturally written in terms of modified Bessel functions, as appropriate for guided modes. For $\gamma^2<0$, the radial dependence is oscillatory, and ordinary Bessel or Hankel functions describe radiating cylindrical waves. In the formulas below, we keep a compact notation in terms of $\kappa_\pm$ and state explicitly which branch is used in each physical regime.

The coupled system~\eqref{eq:coupled_system} is the central result of our formal derivation. It describes how a generic cylindrically symmetric source, $J^\mu(\rho)$, excites electromagnetic potentials in a medium with uniform torsion. As discussed in Appendix~\ref{app:axial_projection}, Eq.~\eqref{eq:coupled_system} isolates the torsional axial mixing within a scalar-envelope transverse approximation; additional vector-cylindrical terms can be incorporated in a full waveguide treatment. The terms proportional to $\Omega$ couple the radial and azimuthal components. Consequently, the eigenmodes are not purely radial or azimuthal; they are circular combinations of the transverse amplitudes. This coupling is the mathematical origin of the polarization rotation and of the direction-dependent response discussed below.

\section{Exact Solution for a Cylindrical Shell Source}
\label{sec:shell_source_solution}

Having established the general wave equations~\eqref{eq:coupled_system}, we now seek an exact analytical solution for a specific, physically illuminating source. We consider an idealized current source that is stationary, purely azimuthal, and infinitesimally thin, localized on a cylindrical shell of radius $R$:
\begin{equation}
    J^{\hat\rho}(\rho)=0,\qquad
    J^{\hat\phi}(\rho)=I_{\phi}\,\delta(\rho-R),\qquad
    J^{\hat z}(\rho)=0 .
    \label{eq:current_profile}
\end{equation}
Here hatted indices denote physical components in the local orthonormal cylindrical frame, and the delta function is a one-dimensional radial distribution normalized as $\int d\rho\,\delta(\rho-R)=1$. Operationally, $I_\phi$ is defined by $\int_{R-\epsilon}^{R+\epsilon}J^{\hat\phi}(\rho)\,d\rho=I_\phi$; it is the integrated azimuthal current amplitude of the thin shell per unit axial length in this effective two-dimensional reduction. In coordinate components, one has $J^\phi=J^{\hat\phi}/\rho$. In the radial equations below, we use the physical transverse component and suppress the hat for notational compactness. This configuration is a standard toy model for sources like solenoidal coils or ring currents in particle accelerators.

The source in Eq.~\eqref{eq:current_profile} is axially symmetric in the sense that $\partial_\phi J^\mu=0$ and translationally invariant along $z$. It is, however, not mirror-symmetric: it carries a definite sense of circulation around the $z$ axis. Thus, the relevant property is not a quadrupolar anisotropy in the transverse intensity profile, but a pseudoscalar handedness associated with the azimuthal current. This distinction is important. The scalar moments of an ideal ring may be rotationally invariant, while the circulation changes sign under parity and couples naturally to the axial torsion of the screw-dislocation background.

This chiral source structure is precisely what makes the torsion-field coupling observable in the present model. The term $K^{\nu}{}_{\mu\lambda}F^{\mu\lambda}$ contains components proportional to $\Omega$ that couple radial and azimuthal field amplitudes. A purely scalar charge distribution or an incoherent current distribution with no net circulation would average out this handed coupling at leading order. The shell current should therefore be viewed as the simplest analytically tractable source that preserves cylindrical symmetry while retaining the handedness needed to excite torsion-induced polarization rotation.

In the homogeneous regions where $\rho \neq R$, the right-hand sides of Eq.~\eqref{eq:coupled_system} are zero. To solve this homogeneous system, it is advantageous to switch to a basis that diagonalizes the coupling matrix. We thus introduce the circular polarization combinations:
\begin{equation}
    a_{\pm}(\rho)\equiv a_{\rho}(\rho)\pm i\,a_{\phi}(\rho).
    \label{eq:polarization_modes}
\end{equation}
These combinations correspond to right-handed ($a_+$) and left-handed ($a_-$) circularly polarized components. As explicitly shown in Appendix~\ref{app:decoupling}, this change of basis transforms the coupled system into two independent, decoupled differential equations:
\begin{equation}
    \Bigl[\frac{\dd^2}{\dd\rho^2}+\frac{1}{\rho}\frac{\dd}{\dd\rho} - \frac{m^{2}}{\rho^{2}}-\kappa_{\pm}^{2}\Bigr]a_{\pm}(\rho)=0.
    \label{eq:bessel_equation}
\end{equation}
This is the Bessel differential equation. Crucially, the two circular polarizations feel different effective radial wavenumbers, $\kappa_\pm$:
\begin{equation}
    \kappa_{\pm}^{2} \equiv \gamma^{2} \mp 2k_{z}\Omega.
    \label{eq:kappa_pm}
\end{equation}
This difference, $\kappa_+^2 \neq \kappa_-^2$, is the mathematical signature of circular birefringence induced by the background torsion.

The general solutions to Eq.~\eqref{eq:bessel_equation} depend on the radial branch.  For radiating modes, $\kappa_\pm^2<0$ after the usual analytic continuation to a real transverse wavenumber, the fields are written in terms of ordinary Bessel and outgoing Hankel functions.  For guided modes, which are the focus of the numerical examples below, it is convenient to set
\begin{equation}
    \chi_\pm^2\equiv \kappa_\pm^2>0,
\end{equation}
so that the regular and decaying solutions are the modified Bessel functions $I_m$ and $K_m$.  The guided-mode solution is therefore
\begin{equation}
    a_{\pm}(\rho)=
    \begin{cases}
        A_{\pm}^{\text{in}}\,I_{m}(\chi_{\pm}\rho), & 0<\rho<R,\\[4pt]
        A_{\pm}^{\text{out}}\,K_{m}(\chi_{\pm}\rho), & \rho>R.
    \end{cases}
    \label{eq:modified_bessel_solution}
\end{equation}
For completeness, the corresponding outgoing radiative solution is
\begin{equation}
    a_{\pm}(\rho)=
    \begin{cases}
        B_{\pm}^{\text{in}}\,J_{m}(q_{\pm}\rho), & 0<\rho<R,\\[4pt]
        B_{\pm}^{\text{out}}\,H_{m}^{(1)}(q_{\pm}\rho), & \rho>R,
    \end{cases}
    \qquad q_\pm^2=-\kappa_\pm^2.
    \label{eq:radiative_bessel_solution}
\end{equation}
Equations~\eqref{eq:modified_bessel_solution} and \eqref{eq:radiative_bessel_solution} are two branches of the same analytic solution. The four amplitudes are determined by matching conditions at the source location, $\rho=R$.

The first matching condition is the continuity of the potential, which implies $a_\rho$ and $a_\phi$, and therefore $a_\pm$, must be continuous at $\rho=R$. The second condition arises from the delta-function source itself. By integrating Eq.~\eqref{eq:wave_aphi_final} across an infinitesimal interval around $\rho=R$ with the radial measure already absorbed in the physical shell-current amplitude, we obtain a jump condition for the derivative of $a_\phi$. This convention is detailed in Appendix~\ref{app:jump_condition} and yields:
\begin{equation}
    \bigl[\partial_{\rho}a_{\phi}\bigr]_{R}\;\equiv\;
    \bigl.\partial_{\rho}a_{\phi}\bigr|_{R^{+}}
    -\bigl.\partial_{\rho}a_{\phi}\bigr|_{R^{-}}
    =\mu_{0}I_{\phi}.
    \label{eq:jump_condition}
\end{equation}
Together, the continuity and jump conditions form a system of four linear equations for the four unknown amplitudes, thereby uniquely determining the electromagnetic field generated by the shell source throughout space.

The shell current plays two roles in what follows. First, it provides an explicitly driven problem from which the relative excitation of the circular eigenmodes can be obtained through the matching conditions. Second, once these guided eigenmodes have been excited, their subsequent longitudinal evolution is described by the Jones and Mueller matrices below. Thus, the source problem and the polarization-propagation problem are complementary: the former fixes modal amplitudes, whereas the latter describes the accumulated phase response of a selected guided mode.

\paragraph*{Remark: consistency with source-free analyses.}
In the limit $I_\phi\to0$ the jump condition~\eqref{eq:jump_condition}
disappears, and the shell current no longer fixes the relative amplitudes of the
circular branches. What remains is the homogeneous torsional propagation problem:
each circular component satisfies Eq.~\eqref{eq:bessel_equation} with transverse
wavenumbers $\kappa_\pm^2=\gamma^2\mp2k_z\Omega$. This reproduces the circular
birefringent splitting found in source-free analyses
\cite{Obukhov2003,Itin2010}. Nontrivial guided modes in the absence of the
source require independent waveguide confinement, material boundary conditions,
or another radial trapping mechanism; they are not generated by the artificial
matching surface at $\rho=R$. The driven formulation therefore reduces, when the
current is removed, to the usual unforced torsional branches, while the finite
shell current supplies the jump condition that determines their excitation
amplitudes.

\section{Polarization Conversion and Direction-Dependent Response}
\label{sec:pol_conv_nonrec}

The analytical solution reveals that a torsion background does more than just permit wave propagation; it actively transforms the polarization state of light and produces a direction-dependent response whose reciprocity properties must be interpreted with care.

\subsection{Jones Matrix and Polarization Rotation}

Let us consider the case of a guided wave propagating along the $z$-axis. The Jones formalism provides a powerful tool for tracking the evolution of a polarization state \cite{JOSA.1941.31.493,PRA.2023.108.033511}. A small technical point is important here. Equation~\eqref{eq:kappa_pm} gives the torsion splitting of the transverse radial problem for a chosen longitudinal separation constant $k_z$. In a waveguide, however, the transverse boundary condition fixes a modal eigenvalue, say $\chi_m$, and the allowed longitudinal propagation constants must adjust accordingly. Setting $\kappa_\pm^2=\chi_m^2$ gives
\begin{equation}
    k_\pm^2-\frac{\omega^2}{c^2}\mp 2\Omega k_\pm = \chi_m^2 .
    \label{eq:modal_dispersion_pm}
\end{equation}
If $\beta_0^2\equiv \omega^2/c^2+\chi_m^2$ is the propagation constant in the absence of torsion and $|\Omega|\ll \beta_0$, Eq.~\eqref{eq:modal_dispersion_pm} gives
\begin{equation}
    k_\pm \simeq \beta_0 \pm \Omega,
    \qquad
    \frac{k_+-k_-}{2}\simeq \Omega .
    \label{eq:weak_torsion_kpm}
\end{equation}
Thus, the transverse splitting found in Eq.~\eqref{eq:kappa_pm} is converted, for a fixed guided mode, into a longitudinal circular birefringence. In the circular basis $\bigl(|R\rangle,|L\rangle\bigr)$, propagation over a distance $L$ simply imparts a different phase to each component. The Jones matrix for forward propagation is therefore diagonal in this basis:
\begin{equation}
    \mathsf{J}_{[\rightarrow]}(L)
    =\begin{pmatrix}
        e^{i k_{+} L} & 0\\[4pt]
        0             & e^{i k_{-} L}
    \end{pmatrix}.
    \label{eq:jones_forward}
\end{equation}
To understand what this means for linearly polarized light, we transform this matrix to the linear basis $\bigl(|x\rangle,|y\rangle\bigr)$. The result is a rotation matrix:
\begin{equation}
    \mathsf{J}_{[\rightarrow]}(L)=
    e^{ik_{\text{avg}}L}\!
    \begin{pmatrix}
        \cos(\Delta k\,L) &  \sin(\Delta k\,L)\\[2pt]
       -\sin(\Delta k\,L) & \cos(\Delta k\,L)
    \end{pmatrix},
    \label{eq:jones_linear}
\end{equation}
where $k_{\text{avg}}\equiv\frac{1}{2}(k_{+}+k_{-})$ is the average propagation constant and $\Delta k\equiv\frac{1}{2}(k_{+}-k_{-})$ is the circular birefringence. To first order in the torsion parameter, $|\Delta k| \approx |\Omega|$; the sign depends on the convention used to label the two circular eigenstates. This means that a linearly polarized wave will have its polarization plane rotated by an angle $\Delta\theta = \Delta k \cdot L \approx \Omega L$. This resembles a polarization rotator at the level of the effective Jones matrix, although its round-trip and two-port scattering properties remain those of a reciprocal optically active medium rather than a Faraday isolator.

\subsection{Reciprocity considerations}

The direction dependence of the torsion-induced splitting must be distinguished from true optical isolation. Reversing the propagation direction changes $k_z\to -k_z$, while the geometric handedness of a fixed screw-dislocation background, represented by $\Omega$, is not automatically reversed. At the level of the local modal equation this changes the sign of the product $k_z\Omega$ and interchanges the two circular branches,
\begin{equation}
    \kappa_{\pm}^2(k_z,\Omega) = \kappa_{\mp}^2(-k_z,\Omega).
    \label{eq:branch_exchange}
\end{equation}
Accordingly, the backward Jones matrix for a fixed laboratory handedness can be written as
\begin{align}
    \mathsf{J}_{[\leftarrow]}(L)&=
    e^{ik_{\text{avg}}L}
    \begin{pmatrix}
        \cos(\Delta k L) & -\sin(\Delta k L)\\[2pt]
        \sin(\Delta k L) &  \cos(\Delta k L)
    \end{pmatrix},
    \label{eq:jones_backward}
\end{align}
up to the same convention-dependent sign used in Eq.~\eqref{eq:jones_linear}.
The rotation angle changes sign under reversal of the propagation direction,
a direction-dependent response that might superficially suggest
non-reciprocal rotation. However, the two-port scattering analysis in Sec.~\ref{subsec:smatrix}
demonstrates that the torsional medium is in fact, Lorentz reciprocal: the
round-trip Jones matrix
$\mathsf{J}_{[\leftarrow]}(L)\cdot\mathsf{J}_{[\rightarrow]}(L)
=e^{2ik_{\rm avg}L}\mathbf{I}$
is proportional to the identity, so the polarization state is fully restored
after a round trip. This places the effect in the same symmetry class as
natural optical activity rather than the Faraday effect.
The formal proof and the conditions under which the response can be used for
device applications are given in Sec.~\ref{subsec:smatrix}.

\subsection{Two-Port Scattering Matrix and Formal Reciprocity}
\label{subsec:smatrix}

The individual Jones matrices~\eqref{eq:jones_forward}--\eqref{eq:jones_backward}
describe the field transformation for a single propagation direction.
To assess the reciprocity of the torsional medium as a device, we embed it
in a standard two-port framework~\cite{Yu2009}.

\paragraph*{Setup.}
Consider a torsional section of length $L$ connecting Port~1 ($z=0$) to
Port~2 ($z=L$). Each port supports two polarization modes, so the full
scattering matrix $\mathsf{S}$ is $4\times4$. In the absence of
reflections ($\mathsf{S}_{11}=\mathsf{S}_{22}=\mathbf{0}$), which holds for
a lossless, impedance-matched torsional medium, the off-diagonal blocks
are the only nonzero entries:
\begin{equation}
    \mathsf{S}
    =\begin{pmatrix} \mathbf{0} & \mathsf{S}_{12}\\
                     \mathsf{S}_{21} & \mathbf{0}\end{pmatrix},
    \qquad
    \mathsf{S}_{21}=\mathsf{J}_{[\rightarrow]}(L),
    \quad
    \mathsf{S}_{12}=\mathsf{J}_{[\leftarrow]}(L).
    \label{eq:smatrix_def}
\end{equation}
Here $\mathsf{S}_{21}$ is the $2\times2$ transmission Jones matrix from
Port~1 to Port~2 (forward), and $\mathsf{S}_{12}$ from Port~2 to Port~1
(backward).

\paragraph*{Lorentz reciprocity.}
The Lorentz reciprocity theorem requires~\cite{Jalas2013}
\begin{equation}
    \mathsf{S}_{12}=\mathsf{S}_{21}^{\mathsf{T}},
    \label{eq:lorentz_recip}
\end{equation}
where ${}^{\mathsf{T}}$ denotes matrix transposition.
Using Eqs.~\eqref{eq:jones_linear} and~\eqref{eq:jones_backward},
\begin{equation}
    \mathsf{S}_{21}^{\mathsf{T}}
    =\bigl[e^{ik_{\rm avg}L}\mathsf{R}(+\Omega L)\bigr]^{\mathsf{T}}
    =e^{ik_{\rm avg}L}\mathsf{R}(-\Omega L)
    =\mathsf{S}_{12},
    \label{eq:recip_check}
\end{equation}
where we used $\mathsf{R}(\theta)^{\mathsf{T}}=\mathsf{R}(-\theta)$.
The condition is satisfied identically, confirming Lorentz reciprocity.

\paragraph*{Round-trip response.}
For a wave propagated forward and then retroreflected, the combined Jones
matrix is
\begin{equation}
    \mathsf{J}_{[\leftarrow]}(L)\cdot\mathsf{J}_{[\rightarrow]}(L)
    =e^{2ik_{\rm avg}L}\,\mathsf{R}(-\Omega L)\,\mathsf{R}(+\Omega L)
    =e^{2ik_{\rm avg}L}\,\mathbf{I}.
    \label{eq:roundtrip}
\end{equation}
The net polarization rotation vanishes after a round trip, with only a
common phase accumulated. This exact identity is the hallmark of
\emph{reciprocal optical activity}: the polarization state is fully restored
after reflection, in contrast to a Faraday rotator, for which the round-trip
rotation doubles to $2\theta_{\rm F}\neq0$.

\paragraph*{Distinction from Faraday rotation.}
Table~\ref{tab:faraday_vs_torsion} summarizes the key differences.
In a Faraday medium, the time-reversal-odd external magnetic field keeps the
rotation sense fixed regardless of propagation direction, yielding
$\mathsf{S}_{12}=\mathsf{S}_{21}=\mathsf{R}(\theta_{\rm F})$ and a round-trip
of $\mathsf{R}(2\theta_{\rm F})\neq\mathbf{I}$.
In the torsional medium, the rotation sense is tied to $k_z$, not to an
external bias, so time-reversal symmetry is preserved.

\begin{table}[h]
\caption{Comparison of torsion-induced optical activity (this work)
with Faraday rotation.}
\label{tab:faraday_vs_torsion}
\begin{ruledtabular}
\begin{tabular}{lcc}
Property & Torsion (this work) & Faraday \\
\hline
$\mathsf{S}_{12}$ & $\mathsf{R}(-\Omega L)$ & $\mathsf{R}(\theta_{\rm F})$ \\
$\mathsf{S}_{21}$ & $\mathsf{R}(+\Omega L)$ & $\mathsf{R}(\theta_{\rm F})$ \\
$\mathsf{S}_{12}=\mathsf{S}_{21}^{\mathsf{T}}$? & Yes & No \\
Round-trip & $\mathbf{I}$ (identity) & $\mathsf{R}(2\theta_{\rm F})$ \\
Time-reversal symmetry & Preserved & Broken \\
Isolation with polarizers? & No & Yes \\
\end{tabular}
\end{ruledtabular}
\end{table}

\paragraph*{Implications for device design.}
Equation~\eqref{eq:roundtrip} establishes that geometric torsion alone cannot
provide optical isolation. Achieving isolator-grade non-reciprocity requires
at least one additional element that breaks time-reversal symmetry, such as
a magneto-optical component, a nonlinear crystal, or a temporal modulator~\cite{Yu2009,Raghu2008}.
The torsional section can, however, serve as a precise, magnet-free
polarization rotator within a larger non-reciprocal circuit, its rotation
angle $\Omega L$ tunable by controlling the defect density.

\paragraph*{Sign convention for polarization rotation.}
Table~\ref{tab:sign_convention} collects the sign of the rotation angle
$\Delta\Theta=\Delta k\cdot L\approx\Omega L$ for all combinations of
propagation direction and Burgers-vector orientation.
In the Jones-matrix convention of Eq.~\eqref{eq:jones_linear}, positive
$\Delta\Theta$ means the polarization vector rotates from $\hat{x}$ toward
$-\hat{y}$ (clockwise in the $xy$ plane when viewed from the $+z$ direction).
The table encodes three key design rules: (i) reversing the dislocation
handedness ($b\to-b$, i.e.\ $\Omega\to-\Omega$) reverses the rotation;
(ii) reversing the propagation direction also reverses the rotation (reciprocal
medium, consistent with Eq.~\eqref{eq:roundtrip}); (iii) the round-trip
rotation (rows~1+3 or 2+4) sums to zero.

\begin{table}[h]
\caption{Sign of the polarization rotation angle
$\Delta\Theta\approx\Omega L$ for all combinations of propagation direction
and Burgers-vector orientation. Positive $\Delta\Theta$ denotes clockwise
rotation viewed from the $+z$ direction in the convention of
Eq.~\eqref{eq:jones_linear}. The round-trip of any pair of rows with the
same $\Omega$ sign but opposite propagation directions gives
$\Delta\Theta_{\rm total}=0$, confirming reciprocity.}
\label{tab:sign_convention}
\begin{ruledtabular}
\begin{tabular}{cccc}
Propagation & Burgers chirality & $\Omega=b\sigma/2$ & $\Delta\Theta$ \\
\hline
$+z$ & Right-handed ($b>0$) & $+|\Omega|$ & $+|\Omega|L$ \\
$+z$ & Left-handed  ($b<0$) & $-|\Omega|$ & $-|\Omega|L$ \\
$-z$ & Right-handed ($b>0$) & $+|\Omega|$ & $-|\Omega|L$ \\
$-z$ & Left-handed  ($b<0$) & $-|\Omega|$ & $+|\Omega|L$ \\
\end{tabular}
\end{ruledtabular}
\end{table}

\subsection{Stokes Parameter Representation and the Poincar\'{e} Sphere}

While the Jones formalism is powerful for tracking the complex electric field of fully polarized light, the Stokes parameters offer a more general and experimentally accessible description \cite{kliger2012polarized,shieh2009ofdm}. They characterize the polarization state using four real, measurable intensity values, allowing them to describe partially polarized or unpolarized light as well. For a plane wave, the Stokes parameters are defined as:
\begin{subequations}
\label{eq:stokes_def}
\begin{align}
    S_0 &= \langle |E_x|^2 + |E_y|^2 \rangle, \\
    S_1 &= \langle |E_x|^2 - |E_y|^2 \rangle, \\
    S_2 &= \langle 2\operatorname{Re}(E_x^*E_y) \rangle, \\
    S_3 &= \langle 2\operatorname{Im}(E_x^*E_y) \rangle.
\end{align}
\end{subequations}
Here, $S_0$ is the total intensity, while $S_1, S_2, S_3$ describe the preference for horizontal, $+45^\circ$, and right-handed circular polarizations, respectively. For fully polarized light, $S_0^2 = S_1^2 + S_2^2 + S_3^2$, which suggests that the normalized state $(S_1/S_0, S_2/S_0, S_3/S_0)$ can be mapped to a point on a unit sphere, the Poincaré sphere \cite{kliger2012polarized}.

The transformation of the Stokes vector $\bm{S}=(S_0, S_1, S_2, S_3)^{\mathsf{T}}$ by an optical element is described by a $4\times4$ real Mueller matrix. The crucial connection to our model is established by identifying the rotation angle $\Delta\theta$ in the general Mueller matrix for a rotator, using our primary result from torsion-induced birefringence: $\Delta\theta \approx \Omega L$. 

By substituting this result, the Mueller matrix becomes the polarimetric signature of our torsional medium for a propagation length $L$:
\begin{equation}
    \mathcal{M}_{[\rightarrow]}(L) = 
    \begin{pmatrix}
      1 & 0 & 0 & 0 \\
      0 & \cos(2\Omega L) & \sin(2\Omega L) & 0 \\
      0 & -\sin(2\Omega L) & \cos(2\Omega L) & 0 \\
      0 & 0 & 0 & 1
    \end{pmatrix}.
    \label{eq:mueller_matrix}
\end{equation}
The action $\bm{S}_{\text{out}} = \mathcal{M}_{[\rightarrow]}(L) \bm{S}_{\text{in}}$ now explicitly depends on our physical parameter $\Omega$. This provides a powerful and intuitive picture: on the Poincaré sphere, after normalization to equal modal power, the ideal lossless longitudinal-mode evolution is a pure rotation of the polarization state about the $S_3$ axis. The rate of this rotation with distance is directly governed by the torsion strength $\Omega$. Equation~\eqref{eq:mueller_matrix} assumes that the two circular guided eigenmodes are normalized to the same attenuation and overlap. If the radial profiles or losses of the two circular branches differ appreciably, the Mueller matrix acquires dichroic terms, and the trajectory need not remain on the equator. This formalism directly predicts the outcome of polarimetric measurements in the ideal phase-birefringent regime.
\begin{figure}[h!]
\centering
\includegraphics[scale=0.45]{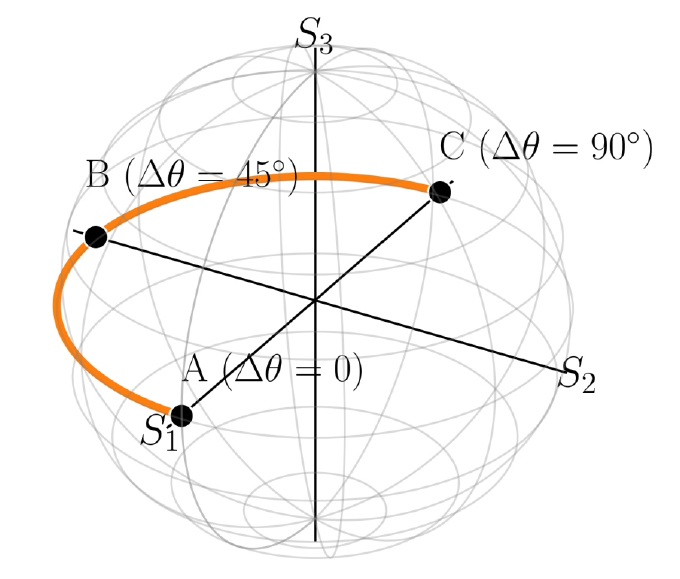}
\caption{Evolution of the polarization state on the Poincaré sphere under the action of torsion. The initial state is a linear horizontal polarization (Point A, on the $S_1$ axis). As the light propagates through the medium, the polarization state traces a circular trajectory (orange line) along the sphere's equator, rotating around the $S_3$ axis. Points B and C represent the states after the plane of polarization has rotated by $45^\circ$ and $90^\circ$, respectively. The fact that the trajectory remains on the equator ($S_3=0$) visually demonstrates the ideal lossless phase-birefringent limit after normalization to equal modal power, in which torsion acts as a pure rotator without induced ellipticity.}
    \label{fig:poincare_sphere}
\end{figure}

The geometric action described by the Mueller matrix in Eq.~\eqref{eq:mueller_matrix} is best visualized on the Poincaré sphere, as depicted in Fig.~\ref{fig:poincare_sphere}. Let us consider an input wave that is linearly polarized along the horizontal axis; its state is represented by Point A on the $S_1$ axis of the sphere. As this wave propagates through the torsional medium over a distance $L$, its polarization state evolves, tracing the circular path shown by the orange trajectory along the sphere's equator. This path represents the precession of the linear polarization plane around the $S_3$ axis. The points B and C mark snapshots of this evolution, corresponding to net polarization rotations of $\Delta\theta = 45^\circ$ and $\Delta\theta = 90^\circ$, respectively. Notably, Point C, where the state vector points along the $-S_1$ axis, corresponds to a 90-degree rotation of the polarization plane and thus represents the peak of 100\% power conversion shown in Fig.~\ref{fig:conversion_efficiency}. This 3D representation provides a powerful and intuitive confirmation of our analytical conclusion: the geometric torsion acts as a pure rotator in the ideal lossless modal limit, modifying the orientation of linear polarization without introducing any circular component. Differential attenuation or unequal radial overlap of the two circular modes would move the Stokes vector away from this equatorial trajectory.

\section{Numerical Results and Discussion}
\label{sec:numerical_results}

To provide a quantitative assessment of the derived effects, we solve the system for a representative set of parameters. We consider light in the telecommunications C-band ($\omega = 2\pi \times 193.4\,\mathrm{THz}$, $\lambda = 1550\,\mathrm{nm}$), propagating in a guided mode ($k_z = 1.01\,\omega/c$), set the azimuthal mode number to $m=0$ (see the discussion in Sec.~\ref{sec:general_formalism}), and take the source current to be located at $R=5\,\mu\mathrm{m}$. For the torsion strength we use $\Omega = 10^4\,\mathrm{m^{-1}}$ as a benchmark value. Since $\Omega=b\sigma/2$, a literal crystal-dislocation realization would require a product $b\sigma\simeq 2\times 10^4\,\mathrm{m^{-1}}$. For a Burgers vector in the nanometer range, this corresponds to very large areal densities, so in the optical context, the parameter should primarily be interpreted as an effective geometric chirality of an engineered metamaterial or defect lattice. This interpretation is consistent with the macroscopic nature of the model: $\Omega$ is the coefficient controlling the observed circular birefringence, not necessarily the density of atomistic dislocations in a natural crystal. The scaling of the results with $\Omega$ is explicit, so smaller effective torsion densities simply increase the required propagation length as $L_{\pi/2}=\pi/(2|\Omega|)$.

\begin{figure}[h!]
    \centering
    \includegraphics[width=0.9\columnwidth]{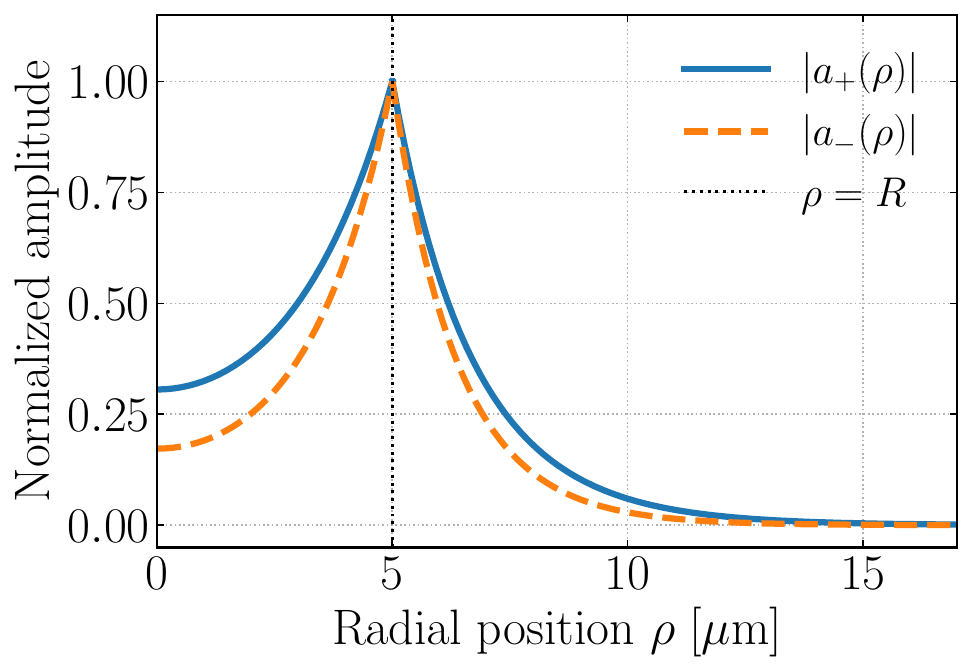}
    \caption{Normalized radial envelopes of the guided circular
    polarization modes, $|a_+(\rho)|$ and $|a_-(\rho)|$, for the
    azimuthal mode $m=0$.
    The profiles are normalized at the shell radius
    $\rho=R=5\,\mu\mathrm{m}$.
    Inside the shell, the regular solution is proportional to
    $I_{0}(\chi_{\pm}\rho)$, while outside it decays as
    $K_{0}(\chi_{\pm}\rho)$.
    The different decay constants
    $\chi_+=4.98\times10^{5}\,\mathrm{m}^{-1}$ and
    $\chi_-=6.42\times10^{5}\,\mathrm{m}^{-1}$ reflect the
    torsion-induced splitting of the two circular branches, with
    a relative splitting $\chi_-/\chi_+-1\approx28.8\%$.
    For $m=0$, the scalar-envelope approximation is exact within
    the transverse truncation
    (see Appendix~\ref{app:axial_projection}).}
    \label{fig:field_profiles}
\end{figure}

Figure~\ref{fig:field_profiles} illustrates the normalized radial envelopes of the guided circular polarization modes, $|a_\pm(\rho)|$, using the modified-Bessel branch of Eq.~\eqref{eq:modified_bessel_solution}. The absolute normalization is fixed by the jump condition in Eq.~\eqref{eq:jump_condition}; the normalization plotted at the shell radius is chosen solely to compare the two circular branches and to emphasize the modal shape. The fields are regular inside the shell, with $a_\pm\propto I_0(\chi_\pm\rho)$, and decay outside as $a_\pm\propto K_0(\chi_\pm\rho)$, characteristic of a guided wave. The different decay constants $\chi_\pm$ reflect the torsion-induced splitting of the two circular branches as predicted by Eq.~\eqref{eq:kappa_pm}; the relative splitting $\chi_-/\chi_+-1\approx28.8\%$ is large enough to be directly observable in the radial profiles.

\begin{figure}[h!]
    \centering
    \includegraphics[width=0.9\columnwidth]{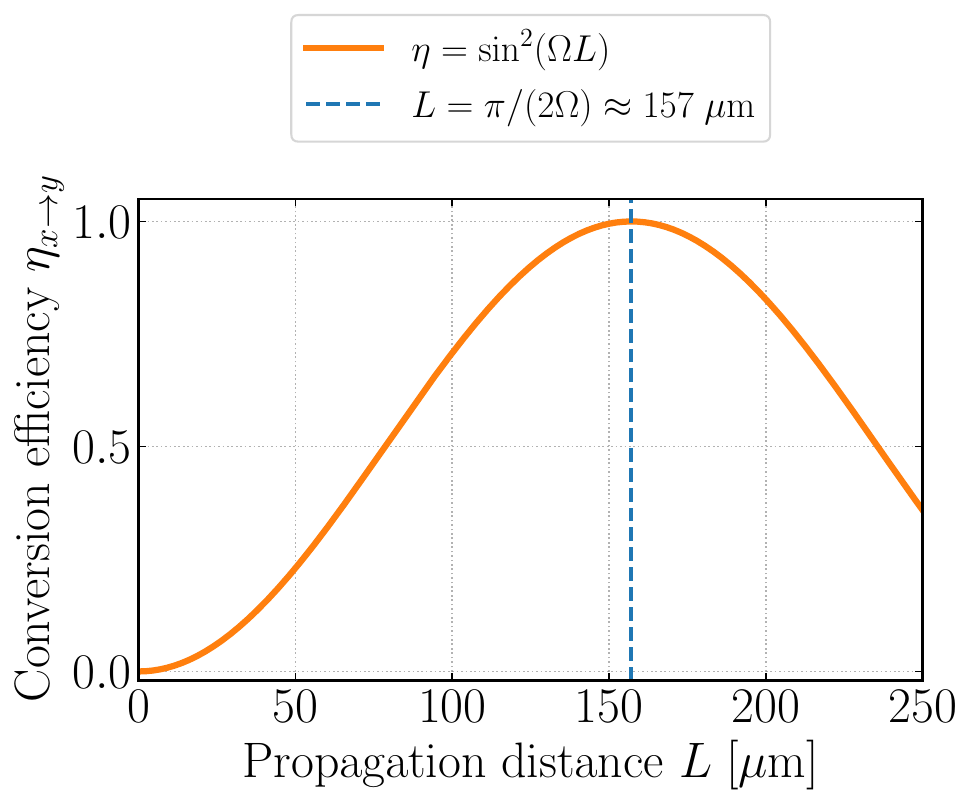}
    \caption{Power converted from the initial $x$-polarization to the orthogonal $y$-polarization as a function of propagation distance $L$ (Eq. \ref{cp}). For a torsion strength of $\Omega = 10^4\,\mathrm{m^{-1}}$, a 90-degree rotation (100\% conversion) is achieved at a distance of $L = \pi/(2\Omega) \approx 157\,\mathrm{\mu m}$. This demonstrates that, in the ideal lossless modal model and for the benchmark effective torsion parameter considered here, geometric torsion can be a highly efficient mechanism for polarization control on micro-photonic scales.}
    \label{fig:conversion_efficiency}
\end{figure}

More striking is the cumulative effect of this birefringence on polarization conversion. For an input wave that is linearly polarized along the $x$-axis, the efficiency of conversion to the orthogonal $y$-polarization after a propagation distance $L$ is given by:
\begin{equation}
    \eta_{x\to y}(L) = \sin^{2}(\Delta k\,L) \approx \sin^2(\Omega L).\label{cp}
\end{equation}
In the torsionless limit \(\Omega\to0\), the two circular branches become
degenerate, \(\kappa_+^2=\kappa_-^2=\gamma^2\), so that
\(\Delta k\to0\) and \(\eta_{x\to y}\to0\), recovering the ordinary non-birefringent propagation limit. Figure~\ref{fig:conversion_efficiency} plots this conversion efficiency as a function of propagation distance for the chosen parameters.

This figure directly visualizes the physical consequences of torsion-induced circular birefringence. The very existence of a non-zero conversion efficiency ($\eta > 0$) is a measurable signature of the effect; in a non-birefringent medium where the circular modes propagate identically ($\Delta k = 0$), the efficiency would remain zero for all distances. The oscillatory, sinusoidal shape of the curve reflects the coherent accumulation of the phase difference between the left- and right-handed circular eigenmodes as they propagate through the medium. The "period" of this energy transfer between polarizations is determined by the magnitude of the birefringence, $\Delta k$, which in our model is directly proportional to the torsion strength, $\Omega$.

The numerical results presented in Fig.~\ref{fig:conversion_efficiency} show that, for the benchmark value of $\Omega$, geometric torsion can act as an efficient mechanism for polarization manipulation. A full 90-degree rotation over a distance of a few hundred micrometers would be highly attractive, in principle, for compact integrated photonic components. Whether the same platform can be used for isolators or circulators requires breaking time-reversal symmetry by an additional element, as established in Sec.~\ref{subsec:smatrix}.

\subsection{Modal Energy Transport and Poynting Analysis}
\label{subsec:poynting}

The Jones and Mueller formalism of the preceding sections tracks the
polarization state, but does not directly expose the energy carried by each
circular eigenmode. We now compute the time-averaged Poynting vector
$\langle\mathbf{S}\rangle = (2\mu_0)^{-1}\operatorname{Re}(\mathbf{E}\times\mathbf{B}^*)$
for the pure $a_\pm$ eigenmodes of Eq.~\eqref{eq:modified_bessel_solution} and
relate the result to the driven excitation amplitudes determined by the shell
source.

\paragraph*{Longitudinal energy flux.}
In radiation gauge ($A_t=A_z=0$, $m=0$), the electromagnetic fields derived
from the ansatz~\eqref{eq:transverse_ansatz_components} satisfy
$\mathbf{E}_\perp = i\omega\mathbf{A}_\perp$ and
$\mathbf{B}_\perp = ik_\pm\hat{z}\times\mathbf{A}_\perp$.
Substituting into the Poynting expression yields the longitudinal flux of
mode $a_\pm$:
\begin{equation}
    \langle S_{\hat{z}}^{(\pm)}\rangle(\rho)
    =
    \frac{\omega\,k_{\pm}}{4\mu_{0}}\,
    \bigl|A_{\pm}^{\rm in}\bigr|^{2}\,
    [f_{\pm}(\rho)]^{2},
    \label{eq:Sz_modes}
\end{equation}
where $f_\pm(\rho)$ is the radial profile of Eq.~\eqref{eq:modified_bessel_solution},
normalized at $\rho=R$. The torsion contributes through two distinct channels:
the different propagation constants $k_\pm\approx\beta_0\pm\Omega$ and the
different profile shapes $f_\pm$ associated with the split transverse wavenumbers
$\chi_\pm$.

\paragraph*{Azimuthal energy circulation.}
For a pure circular branch, $a_-=0$ or $a_+=0$, the definitions
$a_\pm=a_\rho\pm i a_\phi$ imply
$A_{\hat\phi}^{(+)}=-iA_+f_+(\rho)e^{i(k_+z-\omega t)}/2$ and
$A_{\hat\phi}^{(-)}=+iA_-f_-(\rho)e^{i(k_-z-\omega t)}/2$.
The factor $1/2$ in the following expression, therefore, comes from the
conversion between linear and circular transverse components. For $m=0$,
the axial magnetic field is
$B_{\hat z}^{(\pm)}=(1/\rho)\partial_\rho[\rho A_{\hat\phi}^{(\pm)}]$,
and it gives a nonzero azimuthal Poynting component:
\begin{equation}
    \langle S_{\hat{\phi}}^{(\pm)}\rangle(\rho)
    =
    \pm\,
    \frac{\omega\,\bigl|A_{\pm}^{\rm in}\bigr|^{2}}{8\mu_{0}\,\rho}\,
    f_{\pm}(\rho)\,\partial_{\rho}[\rho\,f_{\pm}(\rho)].
    \label{eq:Sphi_modes}
\end{equation}
The opposite signs for $a_+$ and $a_-$ reflect their opposite helicities: the right-handed mode carries a positive (counterclockwise) azimuthal energy
circulation around the shell, while the left-handed mode carries a negative
circulation. Since the numerical benchmark uses $m=0$, this should not be
identified with the canonical orbital angular momentum associated with an azimuthal
phase winding. It is instead a helicity-dependent transverse energy flow, and it
is accessible in principle via transverse near-field scanning or
angular-momentum-sensitive detection.

\paragraph*{Source-driven amplitude asymmetry.}
The matching conditions of Sec.~\ref{sec:shell_source_solution} determine
the driven amplitudes. From the jump condition and the Wronskian identity
$I_0(x)K_1(x)+I_1(x)K_0(x)=1/x$, one finds
\begin{equation}
    A_{+}^{\rm in}=-i\mu_0 I_\phi\,R\,K_0(\chi_+R),
    \;\;\;
    A_{-}^{\rm in}=+i\mu_0 I_\phi\,R\,K_0(\chi_-R).
    \label{eq:amplitudes_driven}
\end{equation}
The amplitude ratio is therefore
\begin{equation}
    \frac{|A_+^{\rm in}|}{|A_-^{\rm in}|}
    =\frac{K_0(\chi_+R)}{K_0(\chi_-R)},
    \label{eq:amplitude_ratio}
\end{equation}
which exceeds unity because $\chi_+<\chi_-$ and $K_0$ is decreasing.
For the benchmark parameters, $K_0(\chi_+R)/K_0(\chi_-R)\approx2.31$, so the
shell source excites the $a_+$ mode more than twice as strongly as $a_-$ in
amplitude. The corresponding modal power ratio
\begin{equation}
    \frac{P_+}{P_-}
    =\frac{k_+}{k_-}
    \left(\frac{K_0(\chi_+R)}{K_0(\chi_-R)}\right)^{\!2}
    \frac{\mathcal{N}_+}{\mathcal{N}_-}
    \approx 6.9,
    \label{eq:power_ratio}
\end{equation}
where $\mathcal{N}_\pm=\int_0^\infty f_\pm^2\rho\,d\rho$ are the modal overlap
integrals. This amplitude asymmetry is the mechanism responsible for the
dichroic terms mentioned in Sec.~\ref{sec:pol_conv_nonrec}: for the driven
problem with unequal modal amplitudes, the Mueller matrix in
Eq.~\eqref{eq:mueller_matrix} acquires off-equatorial corrections proportional
to $(P_+-P_-)/(P_++P_-)$, which would shift the Stokes trajectory
off the equatorial plane. Thus, the ideal Jones--Mueller rotator describes
normalized longitudinal propagation with equal modal power, whereas the
source-driven field contains additional amplitude information fixed by the
shell-current matching conditions.

\begin{figure}[h!]
    \centering
    \includegraphics[width=0.9\columnwidth]{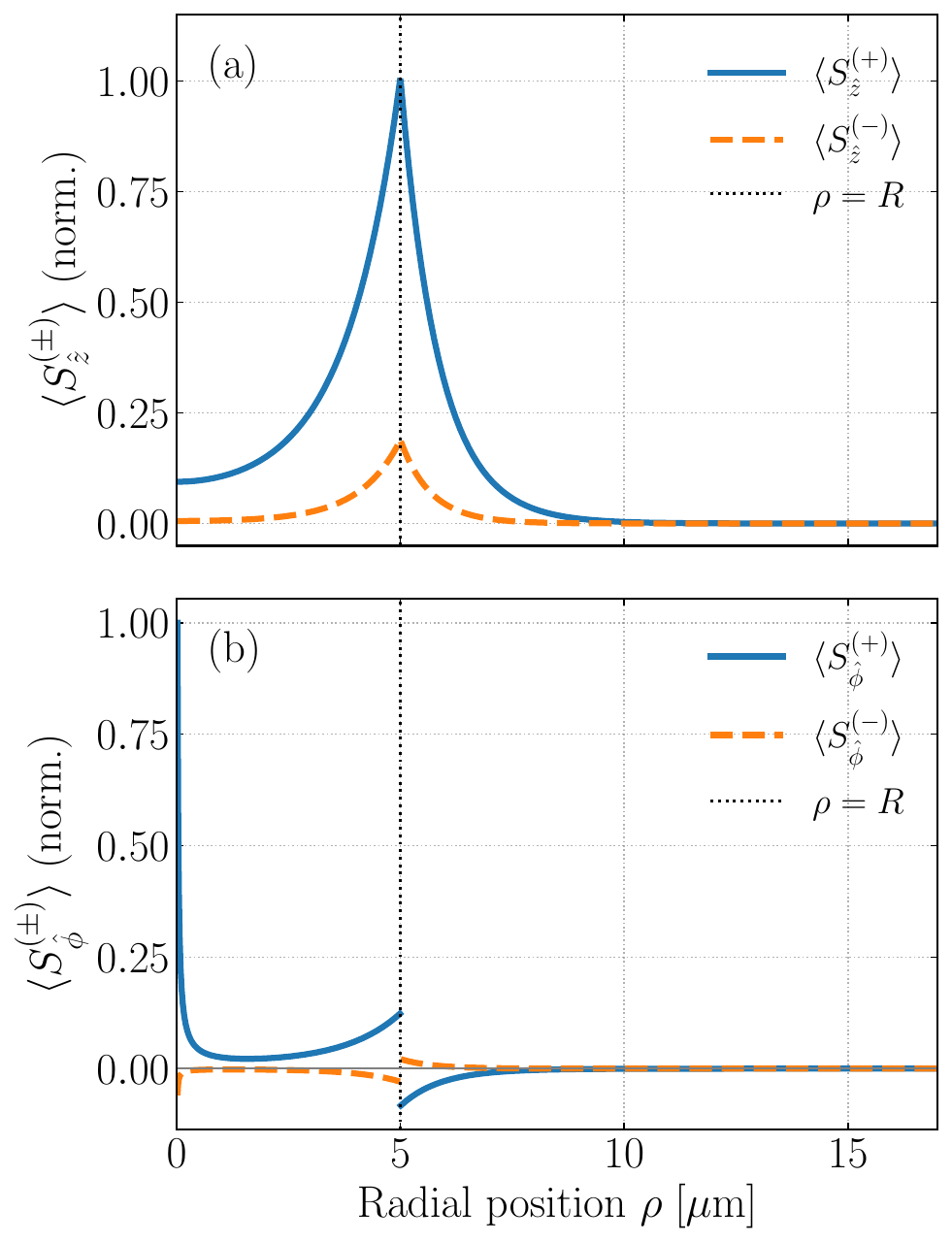}
    \caption{Time-averaged Poynting vector components for the two
    circular eigenmodes, with physically correct driven amplitudes
    from Eq.~\eqref{eq:amplitudes_driven}, normalized to the peak
    value of $\langle S_{\hat{z}}^{(+)}\rangle$.
    (a)~Longitudinal flux $\langle S_{\hat{z}}^{(\pm)}\rangle(\rho)$:
    the $a_+$ mode dominates due to its larger excitation amplitude
    $|A_+^{\rm in}|/|A_-^{\rm in}|=K_0(\chi_+R)/K_0(\chi_-R)\approx2.31$
    and broader radial profile ($\chi_+<\chi_-$).
    (b)~Azimuthal flux $\langle S_{\hat\phi}^{(\pm)}\rangle(\rho)$:
    the opposite signs reflect helicity-dependent azimuthal energy
    circulation in the two eigenmodes, with unequal magnitudes set by the
    torsion-induced amplitude asymmetry of the driven problem.}
    \label{fig:poynting}
\end{figure}

Figure~\ref{fig:poynting} displays the two Poynting components for the benchmark parameters with the correct driven amplitudes.
Panel~(a) shows that $\langle S_{\hat{z}}^{(+)}\rangle$ dominates over
$\langle S_{\hat{z}}^{(-)}\rangle$ across the entire radial domain, reflecting
the combined effect of the Bessel-function amplitude ratio and the broader
profile of the less-confined $a_+$ mode.
Panel~(b) shows the azimuthal flows with opposite signs, their
unequal magnitudes providing a direct measure of the torsion-induced symmetry breaking between the two circular branches.

\section{Conclusions}

In this paper, we have constructed an effective field-theoretical model for the interaction of light with a chiral medium whose chirality emerges from the geometric torsion of its effective spacetime. By systematically including external sources, we have extended previous source-free models, thereby enabling the study of the generation, excitation, and manipulation of light within geometrically chiral optical media.

Our principal conclusions, derived within the effective-action framework, are:
\begin{enumerate}
    \item The dynamics of the electromagnetic potential in the presence of sources and torsion are governed by a system of coupled inhomogeneous wave equations, given by Eq.~\eqref{eq:coupled_system}.
    \item A change to the circular polarization basis diagonalizes the homogeneous part of the system, revealing a torsion-induced circular birefringence. The transverse splitting in Eq.~\eqref{eq:kappa_pm} becomes a longitudinal splitting, Eq.~\eqref{eq:weak_torsion_kpm}, once a guided-mode boundary condition is imposed.
    \item This birefringence leads to a robust and controllable rotation of the linear polarization plane. As shown in our numerical analysis, this effect can be remarkably efficient, allowing, in principle, complete polarization conversion over microscale distances in the ideal lossless modal model for the benchmark effective torsion parameter considered here.
    \item The torsion-induced rotation is direction dependent, as demonstrated by the Jones matrix analysis in Eqs.~\eqref{eq:jones_forward}--\eqref{eq:jones_backward}. The two-port scattering analysis of Sec.~\ref{subsec:smatrix} establishes that the medium is Lorentz reciprocal: the scattering matrix satisfies $\mathsf{S}_{12}=\mathsf{S}_{21}^{\mathsf{T}}$ and the round-trip Jones matrix equals the identity, placing torsion-induced rotation in the same symmetry class as natural optical activity. Achieving magnetic-free optical isolation requires a supplementary element that breaks time-reversal symmetry.
\end{enumerate}

This work lays a solid foundation and opens several promising avenues for future research. On the theoretical front, one could derive the effective coupling from a microscopic action for a specific metamaterial, explore more complex source geometries, investigate finite-dielectric waveguides, and analyze the complete scattering matrix required to distinguish reciprocal optical activity from isolator-grade non-reciprocity. Another natural extension is the dynamic regime in which the torsion parameter $\Omega$ is modulated over time, leading to new parametric effects. On the experimental front, our results provide a roadmap for designing and testing metamaterials that exhibit effective geometric torsion. Finally, the quantization of this field theory could reveal novel phenomena in quantum optics, such as the generation of entangled photon pairs with torsion-mediated correlations.

\appendix

\section{Numerical Residual Check for the Guided Branch}
\label{app:residual_check}
As a consistency check, we verify that the analytical solutions
$I_0(\chi_\pm\rho)/I_0(\chi_\pm R)$ and $K_0(\chi_\pm\rho)/K_0(\chi_\pm R)$
satisfy the decoupled Bessel equations~\eqref{eq:bessel_equation} to
floating-point accuracy. Defining the dimensionless residual $\mathcal{R}_\pm$
via the substitution $x=\chi_\pm\rho$,
\begin{equation}
    \mathcal{R}_\pm(x)
    \equiv
    \frac{\mathrm{d}^2 f_\pm}{\mathrm{d}x^2}
    +\frac{1}{x}\frac{\mathrm{d}f_\pm}{\mathrm{d}x}
    -f_\pm,
    \label{eq:residual_def}
\end{equation}
and evaluating the derivatives via the exact modified-Bessel recurrences
($I_0''=(I_0+I_2)/2$, $K_0''=(K_0+K_2)/2$), one shows analytically that
$\mathcal{R}_\pm\equiv0$ by virtue of the recursion
$I_0(x)-I_2(x)=2I_1(x)/x$. Thus, the normalized modified-Bessel profiles
satisfy the governing Bessel equations identically. In numerical evaluation,
the residual remains at the double-precision roundoff level; no additional
source of algebraic error is introduced by the radial normalization used in
Fig.~\ref{fig:field_profiles}.

\section{Chirality of the Cylindrical--Shell Current}
\label{app:anisotropy_current}

In the main text, the source is idealized as an azimuthal sheet current in the local orthonormal frame,
\begin{equation}
J^{\hat\rho}(\rho)=0,
\quad
J^{\hat\phi}(\rho)=I_{\phi}\,\delta(\rho-R),
\quad
J^{\hat z}=0,
\label{A:eq_profile_cyl}
\end{equation}
where the radial delta is normalized with respect to $d\rho$. The coordinate component is $J^\phi=J^{\hat\phi}/\rho$, but all moment integrals below use the physical vector field. In vector notation, the current is proportional to the azimuthal unit vector,
\begin{equation}
    \mathbf J(\rho,\phi)=I_\phi\,\delta(\rho-R)\,\hat{\boldsymbol\phi}.
    \label{A:eq_profile_vec}
\end{equation}
Using
\begin{equation}
    \hat{\boldsymbol\phi}= -\sin\phi\,\hat{\mathbf x}+\cos\phi\,\hat{\mathbf y}
    =\frac{-y\,\hat{\mathbf x}+x\,\hat{\mathbf y}}{\rho},
\end{equation}
one may write the same source in Cartesian components as
\begin{equation}
    J_x=-I_\phi\sin\phi\,\delta(\rho-R),\qquad
    J_y= I_\phi\cos\phi\,\delta(\rho-R).
    \label{A:eq_profile_cart_components}
\end{equation}

This distribution is rotationally symmetric as a ring: under a rotation about the
$z$ axis maps into itself. Therefore, the usual second-moment tensor in the
transverse plane is proportional to the identity. Strictly speaking, the ideal
radial delta should be understood as the thin-shell limit of a normalized radial
profile $f_\epsilon(\rho-R)$; otherwise products such as $\delta^2(\rho-R)$ are
only formal. With this regularization, the radial factor is a common positive
constant,
\begin{equation}
    \mathcal N_R\equiv \int_0^\infty \rho\, f_\epsilon^2(\rho-R)\,d\rho,
\end{equation}
and the angular dependence gives
\begin{align}
    Q_{xx}&= I_\phi^2\mathcal N_R\int_0^{2\pi}\sin^2\phi\,d\phi
    =\pi I_\phi^2\mathcal N_R,\\
    Q_{yy}&= I_\phi^2\mathcal N_R\int_0^{2\pi}\cos^2\phi\,d\phi
    =\pi I_\phi^2\mathcal N_R,\\
    Q_{xy}&=-I_\phi^2\mathcal N_R\int_0^{2\pi}\sin\phi\cos\phi\,d\phi=0.
\end{align}
Thus $Q_{ij}=\pi I_\phi^2\mathcal N_R\,\delta_{ij}$. The source is not anisotropic in this
second-moment sense.

The physically relevant property is instead its handedness. The ring current has
a definite circulation, quantified by the axial pseudoscalar
\begin{equation}
    \mathcal C_z \equiv \int (\mathbf r\times \mathbf J)_z\,d^2x
    =\int \rho J_\phi\,d^2x.
    \label{A:eq_circulation}
\end{equation}
Substituting Eq.~\eqref{A:eq_profile_vec} yields
\begin{equation}
    \mathcal C_z=I_\phi\int_0^{2\pi}\!d\phi
    \int_0^\infty \rho^2\delta(\rho-R)\,d\rho
    =2\pi I_\phi R^2.
    \label{A:eq_circulation_result}
\end{equation}
This quantity changes sign when the sense of the current is reversed and also
changes sign under parity, while it remains invariant under proper rotations
around the $z$ axis. It is therefore the natural pseudoscalar that couples to
the axial torsion of the screw-dislocation background.

Consequently, the role of the shell source is not to introduce a preferred
linear direction in the transverse plane, but to provide a controlled chiral
excitation with non-zero circulation. This is why it couples efficiently to the
contortion term $K^\nu{}_{\mu\lambda}F^{\mu\lambda}$ in the main text. A source
with vanishing net circulation, such as a purely scalar charge distribution or an
incoherent superposition of counter-circulating currents, would suppress this
handed coupling at leading order.

\section{Projection of the Axial Torsion Term}
\label{app:axial_projection}

Here we show explicitly how the axial Chern--Simons response produces the off-diagonal transverse terms used in Eq.~\eqref{eq:coupled_system}. We work in the local orthonormal frame
$(\hat t,\hat\rho,\hat\phi,\hat z)$, where the metric is locally Minkowskian and the only axial component is $\Theta^{\hat z}=2\Omega$. With the orientation convention of Eq.~\eqref{eq:levicivita_tensor_def}, Eq.~\eqref{eq:K_dual_theta} gives the nonzero transverse contortion components
\begin{equation}
    K^{\hat\rho}{}_{\hat z\hat\phi}=\Omega,
    \qquad
    K^{\hat\phi}{}_{\hat z\hat\rho}=-\Omega,
\end{equation}
together with the components related by antisymmetry in the last two indices. The transverse part of the field equation, therefore, contains
\begin{align}
    K^{\hat\rho}{}_{\hat\mu\hat\lambda}F^{\hat\mu\hat\lambda}
    &=2K^{\hat\rho}{}_{\hat z\hat\phi}F^{\hat z\hat\phi}
      =2\Omega F^{\hat z\hat\phi},\label{eq:proj_rho}\\
    K^{\hat\phi}{}_{\hat\mu\hat\lambda}F^{\hat\mu\hat\lambda}
    &=2K^{\hat\phi}{}_{\hat z\hat\rho}F^{\hat z\hat\rho}
      =-2\Omega F^{\hat z\hat\rho}.\label{eq:proj_phi}
\end{align}
For the transverse ansatz
\begin{equation}
\left\{
\begin{aligned}
    A_{\hat\rho}  &= a_\rho(\rho)e^{i(k_z z-\omega t+m\phi)},\\
    A_{\hat\phi} &= a_\phi(\rho)e^{i(k_z z-\omega t+m\phi)},\\
    A_{\hat z}   &= 0 .
\end{aligned}
\right.
\label{eq:transverse_ansatz_components}
\end{equation}
the longitudinal field-strength components are, to the order retained in the local frame,
\begin{equation}
    F^{\hat z\hat\phi}\simeq i k_z a_\phi,
    \qquad
    F^{\hat z\hat\rho}\simeq i k_z a_\rho .
\end{equation}
Substitution into Eqs.~\eqref{eq:proj_rho}--\eqref{eq:proj_phi} gives precisely
\begin{equation}
    K^{\hat\rho}{}_{\hat\mu\hat\lambda}F^{\hat\mu\hat\lambda}
    =2ik_z\Omega a_\phi,
    \qquad
    K^{\hat\phi}{}_{\hat\mu\hat\lambda}F^{\hat\mu\hat\lambda}
    =-2ik_z\Omega a_\rho .
\end{equation}
These are the off-diagonal terms in Eq.~\eqref{eq:coupled_system}. The diagonal part comes from the projected Maxwell operator. For either transverse component, separation of variables gives
\begin{align}
    &\left[\frac{1}{\rho}\frac{\dd}{\dd\rho}
    \left(\rho\frac{\dd}{\dd\rho}\right)
    -\frac{m^2}{\rho^2}
    -\left(k_z^2-\frac{\omega^2}{c^2}\right)\right]a_i(\rho)
    \notag\\&=
    \left[\frac{\dd^2}{\dd\rho^2}
    +\frac{1}{\rho}\frac{\dd}{\dd\rho}
    -\frac{m^2}{\rho^2}
    -\gamma^2\right]a_i(\rho),
    \qquad i\in\{\rho,\phi\}.
    \label{eq:app_radial_operator}
\end{align}
Terms involving the azimuthal derivative, proportional to $m/\rho$, are therefore included in the Bessel operator after separation of variables. In writing Eq.~\eqref{eq:app_radial_operator} we use the standard scalar-envelope approximation for the transverse physical components. A full vector treatment in cylindrical harmonics generates additional spin-connection (vector Laplacian) terms that mix $a_\rho$ and $a_\phi$ for $m\neq0$; these are discussed below. The displayed torsion contribution is the leading axial mixing term responsible for the circular splitting.

\paragraph*{Validity parameter for general $m$.}
For $m\neq0$, the full cylindrical vector Laplacian acting on the transverse
potential $\mathbf{A}=A_{\hat\rho}\hat\rho+A_{\hat\phi}\hat\phi$ produces
additional geometric cross-coupling terms \cite{Morse1953},
\begin{align}
  \bigl(\nabla^{2}\mathbf{A}\bigr)_{\hat\rho}
    &\supset -\frac{2im}{\rho^{2}}\,a_{\phi},
  \label{eq:vec_lap_rho}\\
  \bigl(\nabla^{2}\mathbf{A}\bigr)_{\hat\phi}
    &\supset +\frac{2im}{\rho^{2}}\,a_{\rho},
  \label{eq:vec_lap_phi}
\end{align}
after applying the azimuthal ansatz $e^{im\phi}$. These terms are distinct
from, and additive to, the torsional Chern--Simons coupling $\pm2ik_z\Omega$.
Their ratio at the characteristic radial scale $\rho\sim R$ defines the
dimensionless control parameter
\begin{equation}
  \xi \;\equiv\; \frac{m}{k_z\,\Omega\,R^{2}}.
  \label{eq:xi_validity}
\end{equation}
For $m=0$, $\xi=0$ and Eq.~\eqref{eq:coupled_system} is \emph{exact} within
the transverse truncation. For $m=1$ with the benchmark parameters of
Sec.~\ref{sec:numerical_results} one would find $\xi\approx0.98$, indicating
that the geometric correction would be comparable to the torsional coupling; a
complete vector treatment would be required in that case. Since the source
in Eq.~\eqref{eq:current_profile} has no $\phi$ dependence and thus excites
only $m=0$, the $m=0$ choice is both physically motivated and theoretically
clean: it eliminates the vector-correction issue entirely, and the condition
$\xi=0$ holds without any constraint on the product $k_z\Omega R^2$.
Modes with $m\neq0$ can still be studied within this framework, but the
scalar-envelope approximation should be supplemented by the full vector
analysis for values of $\xi$ approaching unity.

\section{Decoupling the Homogeneous System}
\label{app:decoupling}

In this appendix, we provide a step-by-step derivation of how the circular polarization combinations $a_\pm(\rho)$ decouple the homogeneous system of wave equations. We begin with the coupled system~\eqref{eq:coupled_system} in the source-free region ($J^\mu=0$):
\begin{subequations}
\begin{align}
    \mathcal{L}[a_\rho] + 2i k_z \Omega a_\phi &= 0, \label{eq:app_hom_rho} \\
    \mathcal{L}[a_\phi] - 2i k_z \Omega a_\rho &= 0, \label{eq:app_hom_phi}
\end{align}
\end{subequations}
where, for brevity, we have defined the linear Bessel differential operator $\mathcal{L}$ as
\begin{equation}
    \mathcal{L} \equiv \frac{\dd^2}{\dd\rho^2} + \frac{1}{\rho}\frac{\dd}{\dd\rho} - \frac{m^2}{\rho^2} - \gamma^2.
\end{equation}
We define the circular polarization fields as $a_\pm = a_\rho \pm i a_\phi$, as in Eq.~\eqref{eq:polarization_modes}.

To find the equation governing $a_+$, we multiply Eq.~\eqref{eq:app_hom_phi} by $i$ and add the resulting equation to Eq.~\eqref{eq:app_hom_rho}:
\begin{equation}
    (\mathcal{L}[a_\rho] + i\mathcal{L}[a_\phi]) + (2ik_z\Omega a_\phi - i(2ik_z\Omega a_\rho)) = 0.
\end{equation}
Because $\mathcal{L}$ is a linear operator, the first term becomes $\mathcal{L}[a_\rho + i a_\phi] = \mathcal{L}[a_+]$. The second term, the coupling term, simplifies as:
\begin{equation}
    2ik_z\Omega a_\phi + 2k_z\Omega a_\rho = 2k_z\Omega(a_\rho + i a_\phi) = 2k_z\Omega a_+.
\end{equation}
Combining these, we obtain a single, decoupled equation for $a_+$:
\begin{equation}
    \mathcal{L}[a_+] + 2k_z\Omega a_+ = 0 \implies (\mathcal{L} + 2k_z\Omega)a_+ = 0.
\end{equation}
Substituting the full expression for $\mathcal{L}$ gives us the final Bessel equation for $a_+$:
\begin{equation}
    \left[ \frac{\dd^2}{\dd\rho^2} + \frac{1}{\rho}\frac{\dd}{\dd\rho} - \frac{m^2}{\rho^2} - (\gamma^2 - 2k_z\Omega) \right]a_+ = 0.
\end{equation}
To find the equation for $a_-$, we follow a similar procedure, but this time we subtract $i$ times Eq.~\eqref{eq:app_hom_phi} from Eq.~\eqref{eq:app_hom_rho}. This yields $(\mathcal{L}[a_\rho] - i\mathcal{L}[a_\phi]) + (2ik_z\Omega a_\phi + i(2ik_z\Omega a_\rho)) = 0$. The first term is $\mathcal{L}[a_-]$, and the coupling term becomes $-2k_z\Omega(a_\rho - i a_\phi) = -2k_z\Omega a_-$. This leads to the decoupled equation for $a_-$:
\begin{equation}
    \left[ \frac{\dd^2}{\dd\rho^2} + \frac{1}{\rho}\frac{\dd}{\dd\rho} - \frac{m^2}{\rho^2} - (\gamma^2 + 2k_z\Omega) \right]a_- = 0.
\end{equation}
This derivation confirms that $a_\pm$ are indeed the eigenmodes of the system, each obeying a distinct Bessel equation with an effective radial wavenumber $\kappa_\pm^2 = \gamma^2 \mp 2k_z\Omega$.

\section{Derivation of the Jump Condition at the Source}
\label{app:jump_condition}

Here, we formally derive the jump condition~\eqref{eq:jump_condition} that the derivative of the potential must satisfy at the location of the delta-function source. This is a standard technique for handling inhomogeneous differential equations with singular sources. We start with the full inhomogeneous equation for $a_\phi$, given by Eq.~\eqref{eq:wave_aphi_final}, with the physical source profile from Eq.~\eqref{eq:current_profile}:
\begin{align}
    &\frac{\dd^2 a_\phi}{\dd\rho^2} + \frac{1}{\rho} \frac{\dd a_\phi}{\dd\rho} - \left( \frac{m^2}{\rho^2} + \gamma^2 \right) a_\phi \nonumber\\
    &\hspace{2cm} - 2i k_z \Omega a_\rho = \mu_0 I_{\phi}\,\delta(\rho-R).
    \label{eq:app_inhom_aphi}
\end{align}
We integrate this entire equation with respect to $\rho$ across an infinitesimal interval straddling the source, from $\rho = R-\epsilon$ to $\rho = R+\epsilon$. Since the singular contribution can only arise from the second derivative, while $(1/\rho)\partial_\rho a_\phi$ is locally integrable across a finite jump in $\partial_\rho a_\phi$, the jump condition can be obtained by integrating the projected radial equation with respect to $d\rho$.
\begin{align}
    &\int_{R-\epsilon}^{R+\epsilon} \frac{\dd^2 a_\phi}{\dd\rho^2} \dd\rho + \int_{R-\epsilon}^{R+\epsilon} \left[ \frac{1}{\rho} \frac{\dd a_\phi}{\dd\rho} - \dots \right] \dd\rho \nonumber \\
    &\hspace{1cm}= \int_{R-\epsilon}^{R+\epsilon} \mu_0 I_{\phi}\,\delta(\rho-R) \dd\rho.
\end{align}
We now analyze each term in the limit where $\epsilon \to 0$:
\begin{enumerate}
    \item The integral of the second derivative gives the jump in the first derivative:
    \begin{equation}
        \lim_{\epsilon\to 0} \int_{R-\epsilon}^{R+\epsilon} \frac{\dd^2 a_\phi}{\dd\rho^2} \dd\rho = \lim_{\epsilon\to 0} \left[ \frac{\dd a_\phi}{\dd\rho} \right]_{R-\epsilon}^{R+\epsilon} = \bigl[\partial_{\rho}a_{\phi}\bigr]_{R}.
    \end{equation}
    \item For any term involving functions that are finite and continuous at $\rho=R$ (such as $a_\phi$ and $a_\rho$, and even $\dd a_\phi/\dd\rho$ itself), their integral over an infinitesimal interval of width $2\epsilon$ will vanish as $\epsilon \to 0$. For instance:
    \begin{equation}
        \lim_{\epsilon\to 0} \int_{R-\epsilon}^{R+\epsilon} \frac{1}{\rho} \frac{\dd a_\phi}{\dd\rho} \dd\rho = \lim_{\epsilon\to 0} (2\epsilon) \times \left( \text{finite value} \right) = 0.
    \end{equation}
    \item The integral over the one-dimensional radial Dirac delta function on the right-hand side is, by definition in our shell-current convention,
    \begin{equation}
        \int_{R-\epsilon}^{R+\epsilon} \mu_0 I_{\phi}\,\delta(\rho-R) \dd\rho = \mu_0 I_\phi.
    \end{equation}
    If one instead writes the source entirely in coordinate components, the factor relating $J^{\hat\phi}$ and $J^\phi$ must be kept together with the cylindrical measure; the same physical jump condition is recovered after expressing the result in terms of the orthonormal current amplitude $I_\phi$.
\end{enumerate}
Equating the surviving terms, we find that only the second-derivative term on the left-hand side contributes, leading directly to the jump condition:
\begin{equation}
    \bigl[\partial_{\rho}a_{\phi}\bigr]_{R} \equiv \left.\frac{\dd a_\phi}{\dd\rho}\right|_{R^{+}} - \left.\frac{\dd a_\phi}{\dd\rho}\right|_{R^{-}} = \mu_0 I_\phi,
\end{equation}
which is the result stated in Eq.~\eqref{eq:jump_condition}.

\section{Orthonormal Frame (Tetrad) Formalism}
\label{app:tetrad}
In curved or non-Riemannian spacetimes, tensor components in a coordinate basis (e.g., $K^\nu{}_{\mu\lambda}$) do not always represent directly measurable physical quantities. It is often necessary to project tensors onto a local orthonormal frame, or tetrad, $e_a^\mu$, where the index $a$ labels the frame vectors (e.g., $a=0,1,2,3$ or $\hat{t},\hat{\rho},\hat{\phi},\hat{z}$) and $\mu$ is the coordinate index. The frame vectors satisfy the orthogonality condition $e_a^\mu e_b^\nu g_{\mu\nu} = \eta_{ab}$, where $\eta_{ab}$ is the Minkowski metric.

For a flat metric in cylindrical coordinates, $ds^2 = -c^2 dt^2 + d\rho^2 + \rho^2 d\phi^2 + dz^2$, the non-zero coordinate-basis components of the tetrad are:
\begin{equation}
    e^t_{\hat{t}} = 1/c, \quad e^\rho_{\hat{\rho}} = 1, \quad e^\phi_{\hat{\phi}} = 1/\rho, \quad e^z_{\hat{z}} = 1.
\end{equation}
The physical components of a tensor are its components in this orthonormal basis. For the contortion tensor, the relationship is:
\begin{equation}
    K^{\hat{a}}{}_{\hat{b}\hat{c}} = e^a_\mu e^\nu_b e^\lambda_c K^\mu{}_{\nu\lambda}.
\end{equation}
The simple form of the contortion tensor presented in Eq.~\eqref{eq:contortion_components}, $K^{\hat\rho}{}_{\hat z\hat\phi}=\Omega$, refers to these physical components. When performing the calculation for the wave equation, one must use the coordinate-basis components, which are found by inverting the transformation. For example:
\begin{equation}
    K^\rho{}_{z\phi} = e^\rho_{\hat{\rho}} e^z_{\hat{z}} e^\phi_{\hat{\phi}} K^{\hat\rho}{}_{\hat z\hat\phi} = (1)(1)(1/\rho) (\Omega) = \Omega/\rho.
\end{equation}
Carefully using these coordinate-basis components in the expansion of Eq.~\eqref{eq:maxwell_sources_torsion} is what correctly yields the factor of $\Omega$ (without any $\rho$ dependence) in the final coupled wave equations~\eqref{eq:coupled_system}. The tetrad formalism is thus a crucial intermediate step that ensures the final physical equations are consistent.

\section*{Acknowledgments}

This work was supported by Conselho Nacional de Desenvolvimento Cient\'{i}fico e Tecnol\'{o}gico (CNPq) (grants 306308/2022-3), Funda\c c\~ao de Amparo \`{a} Pesquisa e ao Desenvolvimento Cient\'{i}fico e Tecnol\'{o}gico do Maranh\~ao (FAPEMA) (grants UNIVERSAL-06395/22), and Coordena\c c\~ao de Aperfei\c coamento de Pessoal de N\'{i}vel Superior (CAPES) - Brazil (Code 001).

%


\end{document}